\documentclass[aps,prl,showpacs,superscriptaddress,floatfix,twocolumn]{revtex4-1}
\usepackage{amsmath,amssymb,graphicx,bm,epsfig}
\usepackage{epstopdf,color}
\usepackage{gensymb}

\begin{document}

\title{Low-Energy Surface States in the Normal State of  $\alpha$-PdBi$_2$ Superconductor}

\author{Hongchul Choi}
\email{chhcl@lanl.gov}
\affiliation{Theoretical Division, Los Alamos National Laboratory, Los Alamos, New Mexico 87545, USA}

\author{Madhab Neupane}
\affiliation{Department of Physics, University of Central Florida, Orlando, Florida 32816, USA}

\author{T. Sasagawa}
\affiliation{Laboratory for Materials and Structures Laboratory, Tokyo Institute of Technology, Kanagawa 226-8503, Japan}

\author{Elbert E. M. Chia}
\affiliation{Division of Physics and Applied Physics, School of Physical and Mathematical Sciences, Nanyang Technological University, 21 Nanyang Link, Singapore 637371, Singapore
}

\author{Jian-Xin Zhu}
\email{jxzhu@lanl.gov}
\affiliation{Theoretical Division, Los Alamos National Laboratory, Los Alamos, New Mexico 87545, USA}
\affiliation{Center for Integrated Nanotechnologies, Los Alamos National Laboratory, Los Alamos, New Mexico 87545, USA}

\begin{abstract}
Topological superconductors as characterized by Majorana surface states have been actively searched for their significance in fundamental science and technological implication. The large spin-orbit coupling in Bi-Pd binaries has stimulated extensive investigations on the topological surface states in these superconducting compounds. Here we report a study of normal-state electronic structure in a centrosymmetric $\alpha$-PdBi$_2$ within density functional theory calculations. By investigating the electronic structure from the bulk to slab geometries in this system, we predict for the first time that  
$\alpha$-PdBi$_2$ can host  orbital-dependent and asymmetric Rashba surface states near the Fermi energy.  This study suggests that $\alpha$-PdBi$_2$ will be a good candidate to explore the relationship between superconductivity and topology in condensed matter physics.
\end{abstract}
	
\date{\today}
\maketitle

{\em Introduction.~}
Recent discovery of quantum materials with gapped bulk but protected metallic surface states has sparked a revolution in condensed matter physics. Among them, although the topological insulators~\cite{MZHasan:2010,XLQi:2011} are now popular, the topological superconductors 
(TSCs)~\cite{AKitaev:2009,APSchnyder:2008, MSato:2009} 
are rare. Nevertheless, the TSCs are of great interest because they can host protected Majorana surface states~\cite{MSato:2009,FWilczek:2009,CIniotakis:2007,AVorontsov:2008,YTanaka:2010,MSato:2010,JAlicea:2012,CWJBeenakker:2013}, 
arising from a nontrivial topology of the bulk Bogoliubov-de-Gennes quasiparticle wavefunctions~\cite{JXZhu:2016}.
The Majorana fermions, that is, hypothetical particles originating from the field of particle physics, have been suggested~\cite{Kitaev01} as a candidate for quantum bits robust against decoherence, motivating intense efforts to materialize the topological superconducting state. 

It is noteworthy to mention that several theoretical proposals have been made to  realize the Majorana states in the condensed matter, from the superfluid phase of  cold atoms to  the interface of the heterostructure with a normal metal, ferromagnetic insulator, superconductor~\cite{Sato:2009,Tanaka:2009,Sau:2010,Tanaka:2012}.  The key ingredients of these systems  are  s-wave superconducting pairing,  Rashba spin-orbit interaction, and  ferromagnetic exchange interaction.  Therefore, searching for the topological superconductivity among heavy element-based compounds is promising to realize Majorana fermion in the real material.  Since the existence of the surface state,  the heavy element-based compound have the strong intrinsic Rashba interaction  and the superconducting pairing to drive a TSC under external magnetic field.

To date, possible topological superconductivity has been suggested in Cu- or Sr-intercalated Bi$_{2}$Se$_3$~\cite{YSHor:2010,MKriener:2011,LAWray:2010,SSasaki:2011,ZLiu:2015,Shruti:2015,MNeupane:2016}, In-doped SnTe~\cite{SSasaki:2012}, highly-pressured Bi$_2$Te$_3$ and Sb$_2$Te$_3$~\cite{JLZhang:2011,JZhu:2013}, 
point-contact induced pressure in Cd$_3$As$_2$~\cite{HWang:2016},
 and hybrid superconductor-semiconductor or -topological insulator heterostructures~\cite{VMourik:2012,MXWang:2012,SYXu:2014}. However,  Pd-Bi 
family~\cite{Matthias63,Zhuravley57a,Zhuravley57b} of superconductors are particularly interesting because both Bi and Pd atoms intrinsically maintain strong spin-orbit coupling (SOC).
 The possibility of the TSC drives the incarnation of the research on Pd-Bi compounds. The noncentrosymmetric $\alpha$-BiPd~\cite{Joshi11} exhibits a superconducting state at $T_c =3.8$ K. If the TSC is realized in $\alpha$-BiPd, the mixture of singlet and triplet pairings is expected due to the absence of inversion center and the strong SOC effect. However, the full gapped superconducting density of states (DOS) was revealed with the scanning tunneling spectroscopy measurement~\cite{Sun15}.
 Also the angle-resolved photoemission (ARPES) experiment demonstrated that the Dirac cone was located at 700 meV below $E_F$~\cite{Neupane15,Thirup16,HMBenia:2016}.  The other member of this Pd-Bi family, PdBi$_2$, preserves the inversion symmetry, and it  has two types of the structures, depending on the annealing temperature.  Recently, $\beta$-PdBi$_2$ had been reported as  a multi or single superconducting 
 gap~\cite{Imai12,Sakano15,Biswas16,Kacmarchik16}.
 The Dirac point  in this structure was detected 2 eV away from $E_F$~\cite{Sakano15}.  Finally, the absence of the TSC was established by the point-contact Andreev reflection spectroscopy~\cite{Che16}.  Although the existence of the topologically protected surface states in $\alpha$-PdBi and $\beta$-PdBi$_2$ has been confirmed, their location of being far away from the Fermi energy negates the possibility of the topological superconducting behavior on the surfaces of these compounds, or in another word, these surface states has no effect on the bulk of the superconductor~\cite{Biswas16}.
 
More recently, $\alpha$-PdBi$_2$  has also been investigated by the penetration depth measurement~\cite{Mitra17}.  Although the data seem to suggest that  the superconducting order parameter has a similar pairing symmetry across the Pd-Bi family of superconductors, the much larger penetration depth itself as deduced for the zero-temperature limit prevents one from ruling out the existence of surface states in this material. This ambiguity suggests the need of a detailed  electronic structure  study  even in the normal state of  $\alpha$-PdBi$_2$, which is currently not available.  Here we report the electronic structure of the bulk and slab structure in $\alpha$-PdBi$_2$ and investigate the characteristics of the surface state.  We find $\alpha$-PdBi$_2$ might be a good candidate for the TSC with the existence of low-energy surface states.

\begin{figure}[tb]
\includegraphics[width=1.0\linewidth]{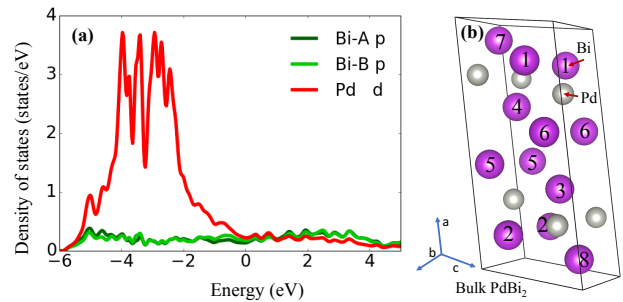}
\caption{(color online)
(a) The density of states (DOS) in the $\alpha$-PdBi$_2$ bulk . The DOS for Bi-A $p$ (dark green), Bi-B $p$ (green) and Pd $d$ (red) electrons were provided between -6 and 5 eV.  (b) The base-centered monoclinic crystal structure of the  $\alpha$-PdBi$_2$ bulk, drawn by VESTA package~\cite{Vesta}. The purple and grey colors represent Bi and Pd atoms. There are two types of Bi atoms such as Bi-A (1-4) and Bi-B (5-8). The appearance of the same number is due to the periodic boundary condition. Hereafter, the energy zero in the DOS denotes the location of the Fermi energy  $E_F$.
}
\label{fig1}
\end{figure}

{\em Method.~}
We employed the density functional theory (DFT) calculation with the full-potential linearized augmented plane wave method implemented in the WIEN2k package~\cite{W2k}. The spin-orbit coupling was considered to describe the heavy elements of Bi and Pd. The Perdew-Burke-Ernzerhof~\cite{PBE} exchange-correlation functional was adopted. The experimental crystal structure was used without the structural relaxation calculation.
The $\alpha$-PdBi$_2$ has a centrosymmetric monoclinic crystal structure of space group $C2/m$. The experimental crystal structure has $a$=12.74$\textrm{\AA}$,  $b$=4.25 $\textrm{\AA}$, $c$=5.665 $\textrm{\AA}$, $\alpha = \gamma =90^{\circ}$, and $\beta=102.58 ^{\circ}$~\cite{Zhuravley57b}. The angle between $a$-axis and $c$-axis is $102.58 ^{\circ}$.  $a$  is two times longer than $b$ and $c$. There are four formula units in the primitive unit cell to have four Pd and eight Bi atoms. Each Bi or Pd atom makes a  mono-atomic single layer stacked along $a$-axis. The crystal structure was visualized in Fig.~\ref{fig1}(b). There are two types of Bi atoms such as Bi-A and Bi-B atoms with the same local symmetry. 1-4 atoms and 5-8 atoms in Fig.~\ref{fig1}(b) correspond Bi-A and Bi-B, respectively. The inversion symmetry is maintained in (1, 2), (3, 4), (5, 6), and (7, 8) pairs.

{\em Results and discussions.~}
At the first stage, we performed the DFT calculation of the bulk PdBi$_2$ to comprehend the underlying electronic property. The valence states of  Pd are made up with $4d^8$ and $5s^2$. The valence state of Bi is expected to have $6p^3$. But, the occupancy of  Pd $4d$ and Bi $6p$ states are computed to show about 8.25 and 1.15 electrons, respectively. The state of Pd $5s$ is almost empty. Around two electrons for each Bi and Pd ions are not included within each muffin radius. Missing electrons would be distributed in the interstitial region to indicate the covalent bond.  
Also, we investigated the DOS in the bulk $\alpha$-PdBi$_2$. Figure \ref{fig1}(a) provides the DOS for Bi-A  $p$, Bi-B $p$, and Pd $d$ states.  The DOSs of Bi-A $p$ and Bi-B $p$ show uniform distribution in this energy range together. Pd $d$ states are mostly occupied to show the several manifestive peaks between -4 and -2 eV.  The valence states are well delocalized to show a metallic feature. Pd $d$ and Bi $p$ states show strong hybridization with the similarity of peaks shown in their DOS. The strong SOC might be associated with the topological surface state through the hybridization.

\begin{figure*}[tb]
	\includegraphics[width=1.0\linewidth]{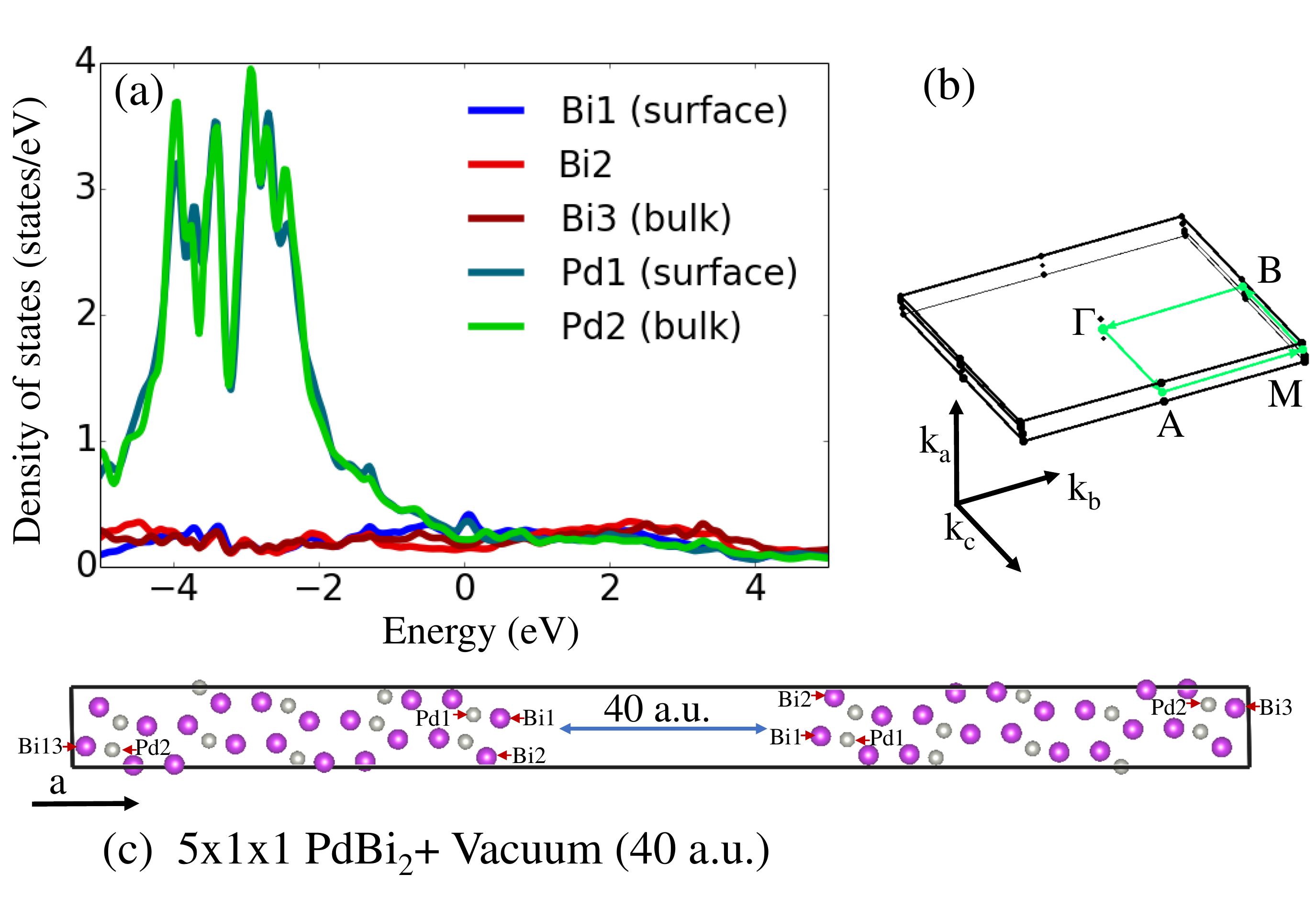}
	\caption{(color online)
		(a) The total  density of states of the selected atoms  in the  $5\times1\times1$ $\alpha$-PdBi$_2$ slab. The used atoms are indicated in Fig.~\ref{fig2}(c). (b) Schematic drawing of the surface  Brillouin zone drawn by Xcrysden package~\cite{{Kokaji03}}. High-symmetric points are marked to be used in the band structure calculation. Due to the elongation along a-axis in the slab structure, the suppression of the Brillouin zone along $k_a$ is shown. To catch the surface states, the $k$-points in the $k_a$=0 plane are chosen. (c) The crystal structure of the  $5\times1\times1$ $\alpha$-PdBi$_2$ slab with 40 a.u., drawn by  the VESTA package.
	}
	\label{fig2}
\end{figure*}
We examined the surface state in the slab geometry. Since the bulk structure has the alternating stack of single Bi and Pd layers along $a$ axis, the bulk structure is elongated five times to become a $5\times1\times1$ unit cell.  The vacuum of 40 a.u. (i.e., Bohr radius) was inserted into the $5\times1\times1$ to construct a slab structure. The slab structure is presented in Fig. \ref{fig2}(c).  The inversion symmetry was maintained to carry out an efficient slab calculation. However, the global inversion symmetry does not affect the numerical result but is associated with computational cost. It is noteworthy that both interfaces are Bi1 layers. 

We investigated how the existence of the interface would affect on the electronic structure. Figure~\ref{fig2}(a) shows the DOS for Bi1-3 and Pd1-2 atoms as marked in Fig. \ref{fig2}(c). The DOS of Bi1 and Pd1  provides the additional high peak at  the Fermi energy. The DOS of Bi2 does not show such a peak around $E_F$.  Therefore, the surface state disappears drastically as going inside into the bulk. Geometrically, the distance between Bi1 and Pd1 is also shorter than that between Bi1 and Bi2. As such, the hybridization between Bi1 and Pd1 might be stronger than that between Bi1 and Bi2. The DOS of Bi3 and Pd2 located deep inside from the surface show the bulk-like DOS. We note that the DOS of Bi1-3 located around -4.5 and 3 eV shows some deviation. This difference might be attributed to modified bonding and anti-bonding states arising from the surface states. Interestingly, the DOS of the surface states is higher than that of the bulk states at $E_F$. When  $\alpha$-PdBi$_2$ enters the superconducting state, the surface states could affect the superconducting pairing, which is in striking contrast to the cases of $\alpha$-PdBi and $\beta$-PdBi$_2$, where the surface states are too far away from the Fermi energy to have any effect on the superconducting pairing. 

\begin{figure*}[tb]
	\includegraphics[width=1.0\linewidth]{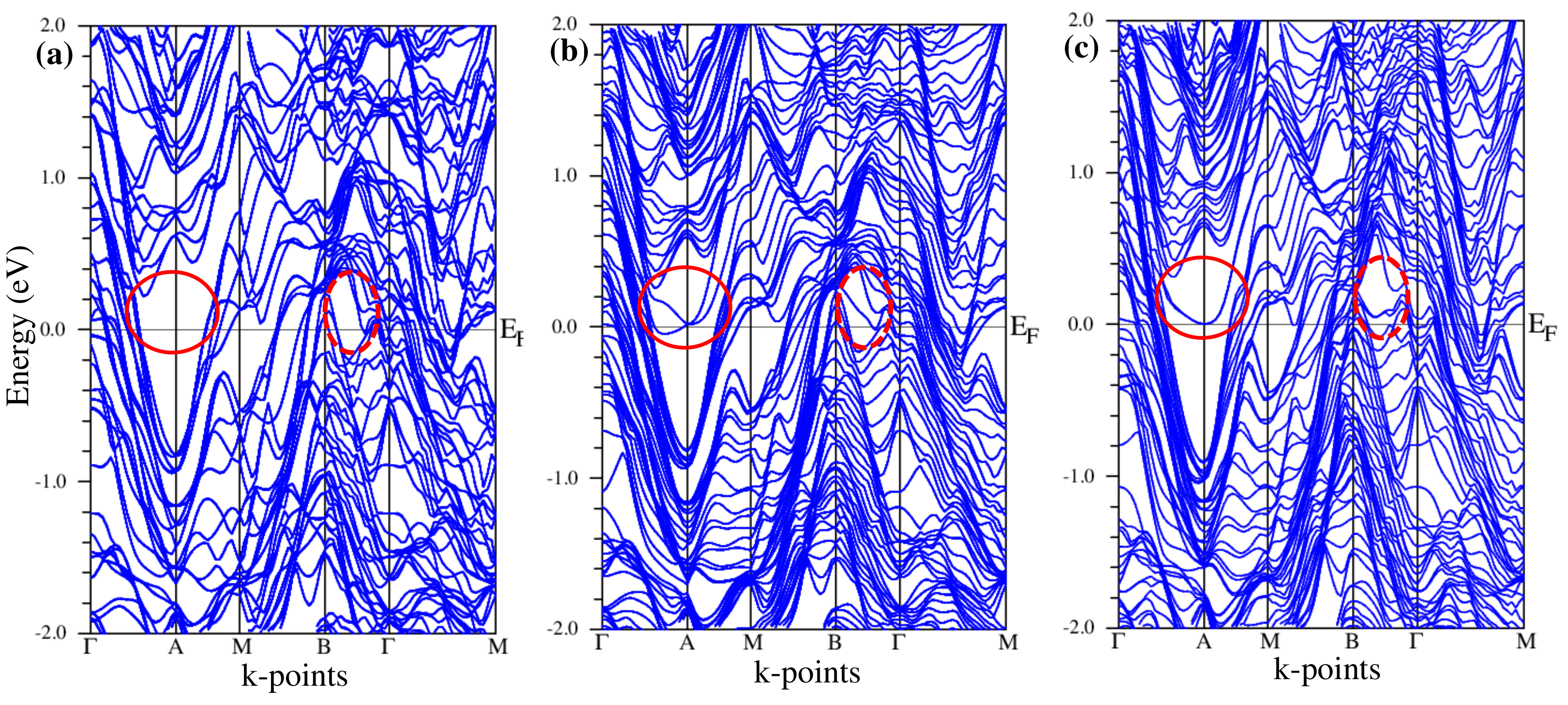}
	\caption{(color online)
	The band structure of the $5\times 1\times1$  $\alpha$-PdBi$_2$ supercell  without  (a)  and with (b) the vacuum.  The slab band structure without the spin-orbit coupling is also provided in the panel (c). The red line and dots are guided to the place we want to discuss.
	}
	\label{fig3}
\end{figure*}
The band structure for the $5\times1\times1$ supercell without and with the vacuum were given in Fig.~\ref{fig3}(a) and (b), respectively. The insulating gap of around 1.4 eV at A point of the wavevector in the Brillouin zone corresponding to  the $5\times 1\times1$ supercell is closed due to the new surface states. Interestingly, the center of the surface bands is located in the proximity of $E_F$.   This surface state is not associated with the global inversion symmetry of the slab structure, but with the local breakdown of the inversion symmetry at the interface between $\alpha$-PdBi$_2$ and vacuum. This assumption is confirmed by the two slab geometries with and without the inversion symmetry to produce the same surface states. Since both interfaces of the slab structure have the same local symmetry, the surface state has a two-fold degeneracy. Besides, the red circles marked around the B and $\Gamma$ points of the wavevector shows the additional clear surface effect around $E_F$.

To understand the SOC effect on the surface state, the slab band structure with and without SO were compared to each other. The band structure without SOC in Fig.~\ref{fig3}(c) shows a parabolic degenerated surface state with one local minimum at the A point. The SOC drives the momentum-dependent splitting (Rashba splitting) of the degenerated surface state to induce several local minimums away from the A point. These SOC-induced Rashba surface states with their energy so close to the Fermi energy opens up a revenue for the emergence of unconventional pairing symmetry at the surface. 
 
\begin{figure}[tb]
	\includegraphics[width=1.0\linewidth]{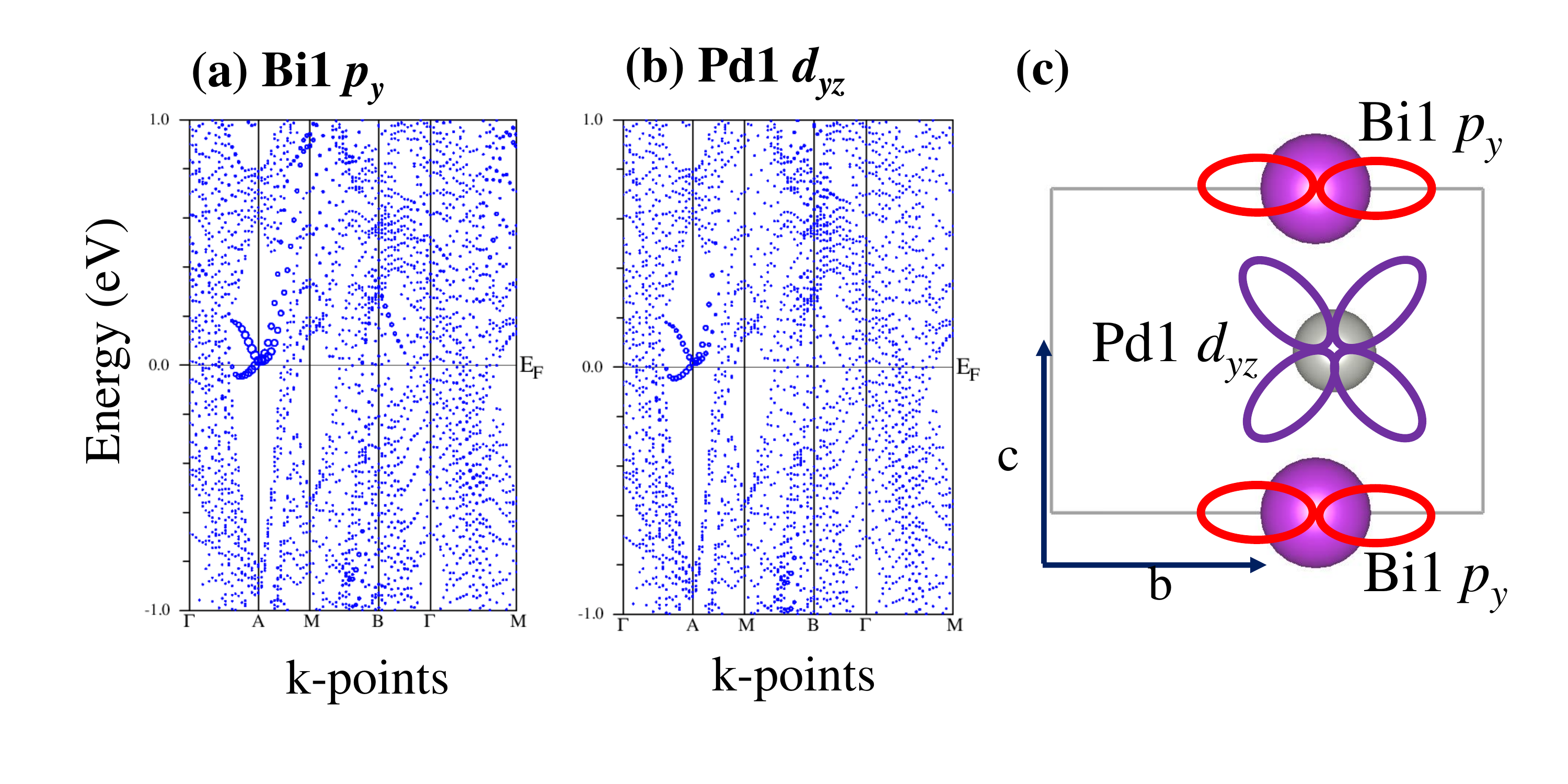}
	\caption{(color online)
	 Angular-momentum-projected slab band structure for (a) Bi1 $p_y$ and (b) Pd1 $d_{yz}$ orbitals. The schematic structure  with Pd1-$d_{yz}$ and Bi1-$p_y$ orbitals  is provided in the panel (c). The other atoms except Pd1 and Bi1 are removed to show the clear relation between Bi1 and Pd1.  
	}
	\label{fig4}
\end{figure}
The characteristic of the edge state at A was analyzed with the angular-momentum-projected orbitals for each atom in the slab geometry. From the DOS shown in Fig.~\ref{fig2}(a), the electronic states derive from Bi1 and Pd1 atoms make up with the surface states. Furthermore, the clarification of the specific angular momentum was investigated. Figure~\ref{fig4}(a) and (b) show that  the surface edge states consist of Bi1 $p_y$ and Pd1 $d_{yz}$ orbitals.  Schematic real-space wavefunction distribution for these two orbitals are displayed in Fig.~\ref{fig4}(c). Bi1 and Pd1 sites have the coordinates of (0.36388,0.61235,0.0) and (0.34137,0.6559,0.5), respectively. The $x$- and $y$-components of their coordinates are very close to each other. They are separated away along $c$-axis.  The relative location between Bi1 $p_y$ and Pd1 $d_{yz}$ suggests the formation of $\pi$-bonding along the $c$-axis. This specific bonding nature might lead to the anisotropic and orbital-dependent surface states.

\begin{figure}[tb]
	\includegraphics[width=1.0\linewidth]{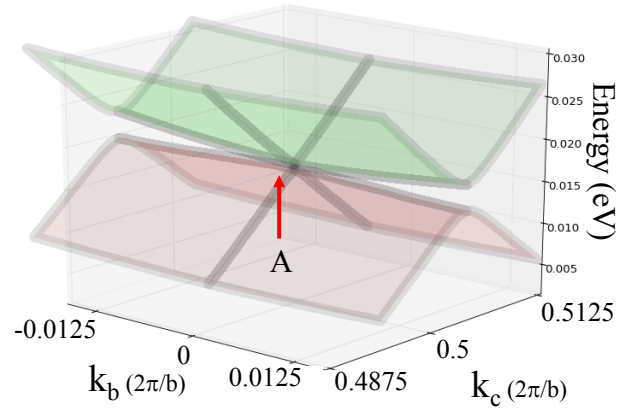}
	\caption{(color online)
      The formation of the Dirac cone at the A-point of the Brillouin zone. The eigenvalues of the two surface states are plotted around the A point  in the $k_b$-$k_c$ plane.
	}
	\label{fig5}
\end{figure}
The detailed topology of the edge states at the $A$ point was examined with the two-dimensional eigenvalues as a function of $k_b$ and $k_c$ around the $A$ point.  Figure~\ref{fig5} shows clearly the existence of the Dirac-cone at the $A$ point. The energy of the Dirac point is 0.017 eV above $E_F$. The slight tuning of the chemical potential could enable the Dirac cone to be established at $E_F$. The dispersion along the $k_c$-direction is much steeper than that along the $k_b$-direction.  This asymmetric Dirac cone is associated with the $\pi$ bonding along the $c$-axis. The asymmetric geometry of the surface states suggests the momentum dependent effective mass of the Dirac quasiparticles. Future ARPES experiments should be able to detect this asymmetric feature of the Dirac cone  in $\alpha$-PdBi$_2$. We would like to note that although there is no theoretical formalism to directly calculate the topology nature of the bulk $\alpha$-PdBi$_2$ metal, the fact that the contribution to the Dirac-like surface states is mainly from the Bi-$p$ orbitals seems to suggest that the bulk electronic structure should be topologically nontrivial. 

 Fu and Kane~\cite{LFu:2008} proposed that topological superconductivity could be achieved through the creation of a ($p_x + ip_y$)-wave superconducting pairing symmetry on the Dirac-cone type surface states inside the topological insulator band gap, when the topological insulator is in proximity to a conventional $s$-wave superconductor. Here the surface states are buried in the metallic bands. Therefore, whether the Cooper pairing of the surface states in $\alpha$-PdBi$_2$ form a TSC  remains an open question, for which the superconducting pair potential must be incorporated explicitly.  Recently, it has been shown for both doped topological insulators~\cite{PHosur:2011} and iron-based superconductors~\cite{GXu:2016} (both are metallic in the normal state) that the TSC with an $s$-wave pairing symmetry on the surface can be realized in an appropriate range of carrier doping. A similar conclusion can be drawn for $\alpha$-PdBi$_2$.

{\em Summary.~}
Through a systematic analysis of electronic structure of  $\alpha$-PdBi$_2$ within the DFT framework, we have identified for the first time the existence of  low-energy surface states in the proximity of $E_F$ in this compound. The bulk DFT calculation shows the delocalized and covalent valence states. The existence of the surface geometry drives the Rashba splitting of the surface state. These split surface states meet at the A point  to produce a Dirac cone in the Brillouin zone.  The Dirac cone is generated by the bonding between Bi1-$p_y$ and Pd1 $d_{yz}$-orbitals in the interface. We have also shown the Dirac cone exhibits orbital-dependent and anisotropic momentum distribution. Since these Dirac states are close to $E_F$ to be easily tunable, the further study of $\alpha$-PdBi$_2$ could shed new light to the realization of the TSC. Interestingly this material will provide the ideal platform to manipulate Rashba spin-orbit coupling and s-wave superconducting pairing. When it is in contact with a ferromagnetic insulator, which provides a ferromagnetic exchange interaction, a topological superconductor which host  Majorana fermions can be realized~\cite{Tanaka:2012}. Furthermore, the high-resolution ARPES measurement in this compound will be necessary to confirm the Rashba surface states near $E_F$.

{\em Acknowledgments.~}
This work was supported by the U.S.\ DOE Contract No.~DE-AC52-06NA25396 through the
LANL LDRD Program (H.C. \& J.-X.Z.), the start-up fund from University of Central Florida (M.N.), 
a CREST project from Japan Science and Technology Agency (JST) and a Grants-in-Aid for Scientic Research (B) from Japan Society for the Promotion of Science (T.S.), and the Ministry of Education (MOE) Tier 1 Grant (RG13/12 and MOE2015-T2-2-065)  (E.E.M.C.)).  
The work was supported in part by the Center for Integrated Nanotechnologies, a DOE BES user facility, in partnership with the LANL Institutional Computing
Program for computational resources.


\begin{thebibliography}{99}
\bibitem{MZHasan:2010} M.Z. Hasan and C. L. Kane, Rev. Mod. Phys. {\bf 82}, 3045 (2010).

\bibitem{XLQi:2011} X. L. Qi and S. C. Zhang, Rev. Mod. Phys. {\bf 83}, 1057 (2011).

\bibitem{AKitaev:2009} A. Kitaev, AIP Conf. Proc. {\bf 1134}, 22 (2009). 

\bibitem{APSchnyder:2008} A. P. Schnyder, S. Ryu, A. Furusaki, and A. W. W. Ludwig, Phys. Rev. B {\bf 78}, 195125 (2008).

\bibitem{MSato:2009} M. Sato and S. Fujimoto, Phys. Rev. B {\bf 79}, 094504 (2009).

\bibitem{FWilczek:2009} F. Wilczek, Nat. Phys. {\bf 5}, 614 (2009).

\bibitem{CIniotakis:2007} C. Iniotakis {\em et al.}, Phys. Rev. B {\bf 76}, 012501 (2007).

\bibitem{AVorontsov:2008} A. Vorontsov, I. Vekhter, and M. Eschrig, Phys. Rev. Lett. {\bf 101}, 127003 (2008).

\bibitem{YTanaka:2010} Y. Tanaka, Y. Mizuno, T. Yokoyama, K. Yada, and M. Sato, Phys. Rev. Lett. {\bf 105}, 097002 (2010).

\bibitem{MSato:2010} M. Sato and S. Fujimoto, Phys. Rev. Lett. {\bf 105}, 217001 (2010).

\bibitem{JAlicea:2012} J. Alicea, Rep. Prog. Phys. {\bf 75}, 075602 (2012).

\bibitem{CWJBeenakker:2013} C. W. J. Beenakker, Annu. Rev. Condens. Matter Phys. {\bf 4}, 113 (2013).

\bibitem{JXZhu:2016} J.-X. Zhu, {\em Bogoliubov-de Gennes Method and Its Applications} (Springer International, Switzerland, 2016).

\bibitem{Kitaev01} A. Y. Kitaev, Phys. Usp. 44, 131 (2001).

\bibitem{Sato:2009} Masatoshi Sato, Yoshiro Takahashi, and Satoshi Fujimoto, Phys. Rev. Lett. {\bf 103}, 020401 (2009).


\bibitem{Tanaka:2009} Yukio Tanaka,  Takehito Yokoyama,  and Naoto Nagaosa,  Phys. Rev. Lett. {\bf 103}, 107002 (2009).


\bibitem{Sau:2010}Jay D. Sau, Roman M. Lutchyn,  Sumanta Tewari, and S. Das Sarma, Phys. Rev. Lett. {\bf 104}, 040502 (2010)

\bibitem{Tanaka:2012} Ai Yamakage, Yukio Tanaka,  and Naoto Nagaosa,  Phys. Rev. Lett. {\bf 108}, 087003 (2012).

\bibitem{YSHor:2010}  Y. S. Hor {\em et al.},  Phys. Rev. Lett. {\bf 104}, 05701 (2010).

\bibitem{MKriener:2011}  M. Kriener {\em  et al.},  Phys. Rev. Lett. {\bf 106}, 127004 (2011).

\bibitem{LAWray:2010}  L. A. Wray {\em et al.},  Nat. Phys. {\bf 6}, 855 (2010).

\bibitem{SSasaki:2011} S. Sasaki {\em et al.}, Phys. Rev. Lett. {\bf 107}, 217001 (2011).

\bibitem{ZLiu:2015} Z. Liu {\em et al.}, J.Am. Chem. Soc. {\bf 137}, 10512 (2015).

\bibitem{Shruti:2015} Shruti {\em et al.},  arXiv:1505.05394v1 (2015).

\bibitem{MNeupane:2016} M. Neupane {\em et al.}, Sci. Rep. {\bf 6}, 22557 (2016).

\bibitem{SSasaki:2012} S. Sasaki {\em et al.}, Phys. Rev. Lett. {\bf 109}, 217004 (2012). 

\bibitem{JLZhang:2011} J. L. Zhang {\em et al.} Proc. Natl. Acad. Sci. {\bf 108}, 24 (2011).

\bibitem{JZhu:2013} J. Zhu {\em et al.}, Sci. Rep. {\bf 3}, 2016 (2013).

\bibitem{HWang:2016} H. Wang {\em et al.}, Nat. Mat. {\bf 15}, 38 (2016).

\bibitem{VMourik:2012} V. Mourik {\em et al.}, Science {\bf 336}, 1003 (2012).

\bibitem{MXWang:2012}  M.-X. Wang {\em et al.}, 
Science {\bf 336}, 52 (2012).

\bibitem{SYXu:2014} 
 S.-Y. Xu {\em et al.},
 Nat. Phys. {\bf 10}, 943 (2014).

\bibitem{Matthias63}B. T. Matthias, T. H. Geballe, and V. B. Compton, \textit{Rev. Mod. Phys.} {\bf 35}, 1 (1963).

\bibitem{Zhuravley57a} N. N. Zhuravlev,\textit {Zh. Eksp. Teor. Fiz.}  {\bf 32}, 1305 (1957).

\bibitem{Zhuravley57b} N. N. Zhuravlev, \textit{JETP} {\bf 5}, 6 (1957)

\bibitem{Joshi11} B. Joshi, A. Thamizhavel, and S. Ramakrishnan, \textit{Phys. Rev. B} {\bf 84}, 064518 (2011).

\bibitem{Sun15} Z. Sun, M. Enayat, A. Maldonado, C. Lithgow, E. Yelland, D. C. Peets, A. Yaresko, A. P. Schnyder, and P. Wahl, \textit{ Nat Commun.}  {\bf 6}, 6633 (2015).

\bibitem{Thirup16}S. Thirupathaiah, Soumi Ghosh, Rajveer Jha, E. D. L. Rienks, Kapildeb Dolui, V. V. Ravi Kishore, B. B\"{u}chner, Tanmoy Das, V. P. S. Awana, D. D. Sarma, and J. Fink, \textit{Phys. Rev. Lett.} {\bf 117}, 177001 (2016).

\bibitem{Neupane15} Madhab Neupane {\it et al.}, \textit{ Nat Commun.}  {\bf 7}, 13315 (2016).

\bibitem{HMBenia:2016} H. M. Benia {\em et al.}, Phys. Rev. B {\bf 94}, 121407(R) (2016).

\bibitem{Sakano15} M. Sakano, K. Okawa, M. Kanou, H. Sanjo, T. Okuda, T. Sasagawa, and K. Ishizaka, \textit{ Nat Commun.}  {\bf 6}, 8595 (2015).

\bibitem{Imai12}Y. Imai, F. Nabeshima, T. Yoshinaka, K. Miyatani, R. Kondo, S. Komiya, I. Tsukada, and A. Maeda, \textit{J. Phys. Soc. Jpn.} {\bf 81}, 113708 (2012).

\bibitem{Biswas16} P. K. Biswas, D. G. Mazzone, R. Sibille, E. Pomjakushina, K. Conder, H. Luetkens, C. Baines, J. L. Gavilano, M. Kenzelmann,  A. Amato, and E. Morenzoni, \textit{Phys. Rev. B} {\bf 93}, 220504(R) (2016).

\bibitem{Kacmarchik16} J. Ka$\breve{c}$mar$\breve{c}$$\acute{i}$k, Z. Pribulov$\acute{a}$, T. Samuely, P. Szab$\acute{o}$, V. Cambel, J. $\breve{S}$olt$\acute{y}$s, E. Herrera, H. Suderow, A. Correa-Orellana, D. Prabhakaran, and P. Samuely, \textit{Phys. Rev. B} {\bf 93}, 144502 (2016).

\bibitem{Che16} Liqiang Che, Tian Le, C. Q. Xu, X. Z. Xing, Zhixiang Shi, Xiaofeng Xu, and Xin Lu, \textit{Phys. Rev. B} {\bf 94}, 024519  (20116).

\bibitem{Mitra17}S. Mitra, K. Okawa, S. Kunniniyil Sudheesh, T. Sasagawa, Jian-Xin Zhu, and Elbert E. M. Chia, \textit{Phys. Rev. B} (accepted, 2017).

\bibitem{W2k}P. Blaha, K. Schwarz, G. Madsen, D. Kvasicka, and J. Luitz, WIEN2K, An Augmented Plane Wave + Local Orbitals Program for Calculating Crystal Properties (Technical University of Vienna, Vienna, 2001).

\bibitem{PBE} J. P. Perdew, K. Burke, and M. Ernzerhof, \textit{Phys. Rev. Lett.}  {\bf 77}, 3865 (1996).

\bibitem{Vesta}K. Momma and F. Izumi, "VESTA 3 for three-dimensional visualization of crystal, volumetric and morphology data," J. Appl. Crystallogr., 44, 1272-1276 (2011).

\bibitem{Kokaji03}A. Kokalj, Comp. Mater. Sci., 2003, Vol. 28, p. 155. Code available from http://www.xcrysden.org/.


\bibitem{LFu:2008} L. Fu and C. L. Kane, Phys. Rev. Lett. {\bf 100}, 096407 (2008).

\bibitem{PHosur:2011} P. Hosur, P. Ghaemi, R. S. K. Mong, and A. Vishwanath, Phys. Rev. Lett. {\bf 107}, 097001 (2011).

\bibitem{GXu:2016} G. Xu, B. Lian, P. Tang, X.-L. Qi, and S.-C. Zhang, Phys. Rev. Lett. {\bf 117}, 047001 (2016).
 


\end{thebibliography}
\end{document}